\documentclass[a4paper,fleqn, 10pt]{article}

\usepackage{geometry}

\usepackage[british]{babel}
\usepackage{amsthm, amssymb, mathtools, amsmath}
\usepackage[dvipsnames,svgnames,x11names]{xcolor}
\usepackage{amsbsy}
\usepackage{enumerate}
\usepackage{mdwlist}
\usepackage{bigints}

\usepackage{enumitem}

\usepackage[numbers,sort&compress]{natbib}

\usepackage{todonotes}
\newcommand{\B}[1]{\boldsymbol{#1}}

\usepackage{graphicx}
\usepackage{subfigure}
\usepackage{float}

\usepackage[normalem]{ulem} 

\usepackage[labelfont=bf,font=small,up]{caption}

\usepackage{lscape} 

\usepackage{url, verbatim}

\usepackage{setspace}

\usepackage{etoolbox}
\makeatletter
\patchcmd{\maketitle}{\@fnsymbol}{\@arabic}{}{}  
\makeatother

\title{Individual preventive measures during an epidemic may have negative population-level outcomes}
\date{\today}
\author{Ka Yin Leung\footnotemark[1] \and Frank Ball\footnotemark[2]\and David Sirl\footnotemark[2] \and Tom Britton\footnotemark[1] }

\begin{document}
\maketitle
\footnotetext[1]{Department of Mathematics, Stockholm University, 106 91 Stockholm, Sweden}
\footnotetext[2]{School of Mathematical Sciences, University of Nottingham, University Park, Nottingham NG7 2RD, UK}
\footnotetext[3]{e-mail: kayin.leung@math.su.se, frank.ball@nottingham.ac.uk, david.sirl@nottingham.ac.uk, tom.britton@math.su.se}

\begin{abstract}
The outbreak of an infectious disease in a human population can lead to individuals responding with preventive measures in an attempt to avoid getting infected. This leads to changes in contact patterns. However, as we show in this paper, rational behaviour at the individual level, such as social distancing from infectious contacts, may not always be beneficial for the population as a whole. We use epidemic network models to demonstrate the potential negative consequences at the population level. We take into account the social structure of the population through several network models. As the epidemic evolves, susceptible individuals may distance themselves from their infectious contacts. Some individuals replace their lost social connections by seeking new ties. We show that social distancing can worsen the disease outcome both in the initial phase of an outbreak and the final epidemic size. Moreover, the same negative effect can arise in real-world networks. Our results suggest that one needs to be careful when targeting behavioural changes as they could potentially \emph{worsen} the epidemic outcome. Furthermore, network structure crucially influences the way that individual-level measures impact the epidemic at the population level. These findings highlight the importance of careful analysis of preventive measures in epidemic models. 
\end{abstract}


\section{Introduction}
Mathematical models for the spread of infections have been succesfully used to increase understanding of how epidemics may propogate: what are the most important features to determine the initial epidemic growth, final epidemic size or endemic level? Mathematical models are also useful to evaluate the possible effects on epidemic dynamics of preventive measures. This can guide public health officials to decide what measures could be put in place to reduce or even stop spreading of a disease~\cite{Heesterbeek2015}. 

To prevent or control an epidemic, public health authorities may implement measures by e.g.\  isolating/treating detected infectious cases or starting a vaccination scheme, either before or during the outbreak~\cite{Heesterbeek2015}. In addition, individuals may take their own measures to prevent themselves from getting infected, e.g.\ by wearing face masks, taking hygienic measures such as hand washing, or by socially distancing themselves from infectious contacts. Such individual behaviour has been observed in e.g.\ the recent Ebola outbreak and the 2009 A/H1N1 epidemic~\cite{Rubin2009, Jones2009,Bayham2015, Fast2015, Funk2017}.

In general, it is hard to predict the effect of preventive measures without using models to guide us. Epidemic dynamics are highly nonlinear and therefore preventive measures can lead to counter-intuitive effects. Standard epidemic models assume human behaviour is not influenced by the epidemic and is constant over time. Although it is often recognized that humans do take preventive measures in the course of an epidemic, models that incorporate behavioural dynamics are generally much harder to analyze. Recently, such models have started to receive more attention, and important advances have been made to gain understanding of the effect of different behavioural changes on epidemic dynamics~\cite{Funk2010,Manfredi2013,Funk2015,Verelst2016}. 

A crucial modelling ingredient is the contact pattern in the population as infection is transmitted through contacts between susceptible and infectious individuals. Owing to challenges in their analysis, the majority of models that consider behavioural responses to epidemic dynamics are relatively simple in modelling contact patterns~\cite{Verelst2016}. Often the simplest assumption of homogeneous mixing, or some variant, is made. This assumption implies that any two individuals rarely meet more than once in a large population. To overcome the restriction of the lack of repeated contacts, network epidemic models have been proposed to model human contact patterns. This class of models have received much attention over the last 20 years or so~\cite{Newman2010, Danon2011}. In these models, individuals are socially connected in the network and infection is only possible along connections. Network models are also a natural way to incorporate heterogeneity in the number of connections that individuals in the population have. Throughout this paper, we refer to two individuals that are connected to each other as `neighbours'. Exactly what a neighbour is depends on the social structure under consideration, e.g. one may think of the neighbours as `colleagues' in workplaces or `sexual partners' in sexual networks. 

In the current paper we study a network SIR epidemic with preventive social distancing. We consider the setting where susceptible individuals distance themselves from their neighbours who they find out are infectious, perhaps sometimes simply dropping such connections and other times, in their wish to maintain a certain number of social connections, by seeking new connections (which we refer to as `rewiring'). We study the impact of social distancing on model networks as well as real-world networks. 

We show that rational preventive individual-level behaviour can have counter-intuitive negative population-level consequences. From the perspective of an individual who distances him/herself from an infectious individual, this preventive behaviour is always rational in the sense that it decreases the risk of him/her getting infected during the epidemic outbreak (here `always' means for all rewiring and dropping rates on all networks). However, we also show that having individuals who rewire away from infectious neighbours and possibly replace them with new ties \emph{may be harmful} for the community as a whole. Depending on the network structure of the population, social distancing may in fact \emph{increase} the epidemic threshold parameter from below to above its threshold value, making a large outbreak possible where without social distancing it was not. We also show that social distancing can \emph{increase the final size} of the epidemic. It is important to stress that these features do not hold for all networks. However, we show that there are real-world networks as well as model networks which exhibit these properties. It is difficult to characterize completely when such individual preventive behaviour is harmful, but it tends to happen more easily if: a) the basic reproduction number $R_0$ or the related clique reproduction number $R_*$ (for the baseline setting without social distancing) is large, b) the network has many individuals with low degree and possibly other groups being highly inter-connected, and c) connections are more likely to be rewired than dropped. 

\section{Model}
\subsection{SIR epidemic with social distancing on a network} 
We consider a population in which individuals are socially connected. Two individuals that are connected to each other are referred to as neighbours and contacts are only made between neighbours. The individuals and the connections between them together make up the network structure of the population. The stochastic SIR (susceptible-infectious-recovered) epidemic with social distancing on a network is as follows. Initially, usually one individual is infectious, we call this individual the index case, and all others in the population are susceptible (specific assumptions concerning the index case are given later). An individual that gets infected becomes infectious and remains so for an exponentially distributed time with mean $1/\gamma$. During its infectious period an individual transmits infection at a constant rate $\beta$ independently to each susceptible neighbour. Moreover, a susceptible individual that has an infectious neighbour drops that connection at rate $\omega_d$ and rewires the connection to an individual chosen uniformly at random from the population at rate $\omega_r$. We write $\omega=\omega_r+\omega_d$ for the total social distancing rate. One can think of $\omega_r=\alpha\omega$ and $\omega_d=(1-\alpha)\omega$, where $\omega$ represents the rate at which a susceptible individual finds out that a given neighbour is infectious, and $\alpha=\omega_r/(\omega_r+\omega_d)$ is the probability that the individual wishes to retain his/her number of neighbours. Dropping and rewiring events happen independently between all pairs of susceptible and infectious individuals. The epidemic continues until there is no connected susceptible-infectious pair of individuals. 

Note that the preventive measure of social distancing is always beneficial from the individual perspective. Indeed, a susceptible individual that distances itself from an infectious neighbour avoids the risk of getting infected by that particular individual. In the case that it chooses to replace that social connection (rewiring), and that new neighbour is recovered (and immune), transmission can no longer occur through that connection. If the neighbour is susceptible, transmission through that connection could occur later on in the epidemic. If the neighbour is infectious, then all that has happened from an epidemic point of view is that one infectious neighbour is replaced by another one, and the risk of becoming infected is unchanged. Obviously, the most beneficial option from the point of view of avoiding getting infected is not to replace the connection (corresponding to $\omega_r=0$ and $\omega_d>0$ in the model). This extreme case of dropping connections is always beneficial from both the individual and population perspective and can in fact be analysed mathematically (F Ball, T Britton, KY Leung, D Sirl (2018). An SIR network epidemic model with preventive dropping of edges. \emph{Manuscript in preparation}).

The epidemic with social distancing is studied on two network models as well as two real-world networks. The networks are described in Section~\ref{sec:networks} below. Our results in Section~\ref{sec:results} involve several epidemiological measures for the beginning and the end of the epidemic, these concepts are introduced in Section~\ref{sec:analysis}.


\subsection{The networks}\label{sec:networks}
\subsubsection{Configuration network}\label{sec:configuration}
The configuration model is a well-studied network, both within and without the context of epidemic models~\cite{Bollobas2001,Molloy1995,Newman2001}. The network is constructed by first defining its degree distribution $\{p_d\}$, where $p_d$ is the probability that an individual has exactly $d$ connections. In a population of size $n$, each of the $n$ individuals picks a degree independently from $\{p_d\}$ and attaches that many half-edges to itself. Half-edges are then paired completely at random and the corresponding individuals are connected in the network. By way of this construction, some imperfections may arise, such as self loops or multiple connections between some pairs of individuals. However, such imperfections become sparse in the network as the population size $n\to\infty$ if the degree distribution has finite variance (see e.g.~\cite[Theorem 3.1.2]{Durrett2006}). Therefore, it is safe to remove such imperfections and assume that the configuration network is a simple undirected network with the prescribed degree distribution~\cite{Janson2009}. We denote the mean and variance of the degree distribution $\{p_d\}$ by $\mu_D$ and $\sigma_D^2$, respectively. 

\subsubsection{Clique network}\label{sec:cliques}
The clique-network model~\cite{Ball2009} (also referred to as household-network model when the unit under consideration is interpreted as a household) has two types of connections: global network connections and clique connections. The global network structure is obtained through the configuration network with prescribed degree distribution $\{ p_d\}$. On top of this, the community is partitioned into distinct units (cliques) of size three (see \textit{SI Section~S2} for a discussion on allowing for various clique sizes). The population can be partitioned into cliques by labelling all individuals from 1 to $n$, and letting the first three individuals make up clique 1, the next three individuals make up clique 2, and so on. In the final network, individual 1 is then connected to all individuals he/she is connected to from the construction of the configuration model together with individuals 2 and 3 from the clique construction, and similarly for the other individuals. As with the configuration network, the clique configuration network can be treated as a simple undirected network.

\subsubsection{Real-world networks}\label{sec:realworld}
The real-world networks for our studies are taken from the Stanford large network dataset collection~\cite{snapnets}, where datasets for several different networks are freely available. We considered the `collaboration network for arXiv General Relativity' and the `Facebook social circles network'. Both networks are undirected. The `arXiv General Relativity collaboration' network describes scientific collaborations between authors that submitted papers to the arXiv in the General Relativity and Quantum Cosmology category. Edges between nodes represent two co-authors that have written a paper together. In the `Facebook social circles' network, nodes are survey participants of the social network website Facebook that were using a specific app. Edges between nodes represent the `circles' or `friends lists' of those participants. More details are found in~\cite{snapnets} and~\textit{SI Section~S3.1}.

\subsection{Epidemiological quantities: $\B{R_0}$, $\B{R_*}$, and the final size}\label{sec:analysis}
In general, the social distancing model is challenging to analyze mathematically (see~\cite{Britton2016} for analysis of the beginning of an epidemic on the configuration network). As the network structure depends on the epidemic dynamics, models very soon become intractable. Therefore, in the main text we present the heuristics of our analytical results and refer to \textit{SI} for the mathematical details. In Section~\ref{sec:results} the main focus is on our findings from simulation studies. Here, we present the key epidemiological concepts that are used in Section~\ref{sec:results}.

For the beginning of the epidemic, in the configuration network model we use the basic reproduction number $R_0$ that has the interpretation as the expected number of secondary cases generated by one typical newly infected individual at the beginning of the epidemic. The number $R_0$ is a threshold parameter with threshold value one in the sense that, in the limit as the population size $n\to\infty$, there is a positive probability of a major outbreak (one which infects a strictly positive fraction of the population as $n\to\infty$) if $R_0>1$ and no major outbreak occurs if $R_0\leq1$. Owing to stochastic effects, it is always possible that an epidemic dies out when introduced into a population (with finite size $n$) even when $R_0>1$. Previous work (\cite{Britton2016}; see also \textit{SI Section~S1.2}) showed that the basic reproduction number $R_0$ for the epidemic on the configuration network with social distancing is given by 
\begin{equation}\label{eq:R0_configuration}
R_0=\frac{\beta}{\beta+\omega+\gamma}\left(\mu_D+\frac{\sigma_D^2}{\mu_D}-1\right),
\end{equation}
where $\mu_D+\sigma_D^2/\mu_D-1$ is the expected number of susceptible connections of a typical newly infected individual in the early stages of an epidemic and $\beta/(\beta+\omega+\gamma)$ is the probability of transmitting to such a susceptible individual before he/she recovers or the neighbour drops the connection or rewires away. 

Related to $R_0$ is the clique reproduction number $R_*$ (also referred to as the household reproduction number when the cliques under consideration are households), which is more natural to consider when studying populations with a clique structure. Rather than considering a newly infected individual, one considers a newly infected clique as the unit of interest. The same threshold behaviour holds. The clique reproduction number $R_*$ is derived in~Section~\ref{sec:clique} and \textit{SI~Section S2.1}.

For an epidemic on both the configuration network and the clique network, as population size $n$ tends to infinity, the final fraction $\bar Z_n$ of individuals that ever get infected converges in distribution to random variable $\bar Z$ with two-point distribution: $P(\bar Z=0)=1-P(\bar Z=z)$. In the event of a major outbreak, the limiting final fraction of the population infected by the epidemic is $z$. In general, this constant $z$ is only characterized implicitly, even for the simplest Markovian homogeneously mixing SIR epidemic model. We use the practical definition in our simulation studies in Section~\ref{sec:results} that an epidemic outbreak is major if the final number of infected individuals is more than $10\%$ of the total population size. Furthermore, without loss of generality, we scale time such that the transmission rate is $\beta=1$ per time unit, so all rates are interpreted as relative to the transmission rate $\beta$. More details on the simulation studies are provided in \textit{SI Section S4}. We call the model without social distancing ($\omega_r=0=\omega_d$) the baseline model.

\section{Results}\label{sec:results}
\subsection{The configuration network}
\begin{figure}[tbhp]
\centering
\includegraphics[width=.4\linewidth]{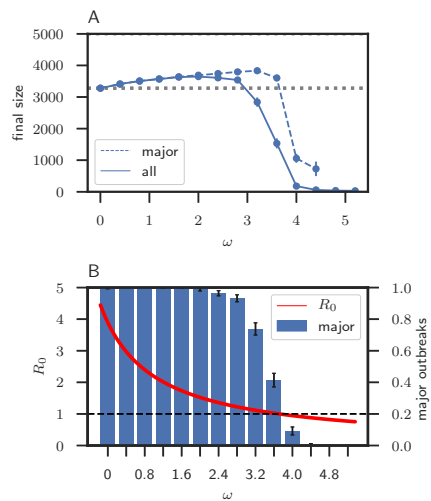}
\caption{Social distancing can lead to an increase in the final size for the configuration network model. (A) Average final size (with 95\% confidence intervals (CI)) over all outbreaks (solid line) and restricted to major outbreaks (dashed line); the dotted horizontal line is at the final size when $\omega=0$, for reference. (B) $R_0$ as a function of social distancing rate $\omega$ (dashed black line at $R_0=1$ indicates the threshold value) and fraction of all outbreaks resulting in major outbreaks (with 95\% CI). Model parameters are as follows. An individual in the population has degree $d$ with $d=0,\ldots,10$ with probability $p_d=c/(d+1)$, $d=0,1,\ldots,10$, with $c=0.331$ the normalization constant. Other parameter values are $\omega_r/\omega_d=9$ and $1/\gamma=10$ time units, total population size 5000, and each epidemic starts with 10 randomly chosen index cases. For each value of $\omega$, 500 epidemics are simulated.}  
\label{fig:config_finalsize}
\end{figure}

Social distancing in the configuration network is always beneficial at the beginning of an epidemic in the sense that it lowers $R_0$. This conclusion follows immediately from expression~\eqref{eq:R0_configuration}. In fact, social distancing can ensure that $R_0$ is reduced below the epidemic threshold value of one, see Fig.~\ref{fig:config_finalsize}B for an example. At the beginning of an epidemic, from the point of view of a susceptible individual, social distancing from an infective neighbour more or less ensures that he/she avoids infection. Indeed, there are only few infectives in the population in that stage of the epidemic. This makes it unlikely for a susceptible individual to encounter another infectious individual at the beginning of the epidemic.

However, social distancing need not be beneficial for the population as a whole. In fact, even though rewiring decreases $R_0$, it can still lead to an increase in the final size. To show analytically that the expected final size can increase with $\omega$ we consider a very specific degree distribution, where individuals have either degree 0 or degree $k$, where $k>2$, i.e.\ $p_0=1-p_k$ (proving things for more general degree distributions seems very hard). We analyze a related model that allows us to derive an asymptotic lower bound for the model of interest. In the related model, we consider an SI infection ($\gamma=0$). Then continuity arguments ensure that our results also hold for an SIR infection with $\gamma>0$ small enough. Individuals act differently depending on their degree. A susceptible individual that tries to rewire to a randomly chosen individual $v$ in the population will not do so if $v$ is of degree $k$. If $v$ is of degree 0, then rewiring takes place as usual, but $v$ is prohibited from transmitting to other individuals. Therefore, the number of infections in the modified model is always less than in the original model (and is equal in the baseline model when there is no social distancing). For this modified model, we can derive an expression for the asymptotic final size, yielding a lower bound for that quantity in our social distancing model, and consequently proves that the final size can increase for small $\omega>0$ and $\gamma>0$. The details of the analysis are found in \textit{SI Section~S1.3}.

Rather than providing details for the analytical results for the final size here, we demonstrate the negative population level effects through simulation studies. We consider the social distancing model on a configuration network with heterogeneous degree distribution in Fig.~\ref{fig:config_finalsize}. Parameter values are such that the basic reproduction number $R_0$ is large in the baseline setting and the majority of the social distancing is done through rewiring rather than dropping. The epidemic is started with 10 index cases (chosen uniformly at random from the population) in order to have most of the simulations resulting in major outbreaks. Additional results showing that social distancing can increase the final size for several other configuration network models are presented in \textit{SI~Section S1.4}. 

Note that the fraction of epidemics that result in major outbreaks decreases with increasing social distancing rates (Fig.~\ref{fig:config_finalsize}B). Once the social distancing rate $\omega$ increases to a level such that the basic reproduction number drops below the epidemic threshold value of one (Fig.~\ref{fig:config_finalsize}B), mostly minor outbreaks will occur. Finally, we note that deviations from the average final size are generally small (also compared to the total population size of 5000), especially when conditioning on the occurrence of a major outbreak. 

\subsection{The clique network}\label{sec:clique}
\begin{figure*}[tbhp]
\centering
\includegraphics[width=.8\linewidth]{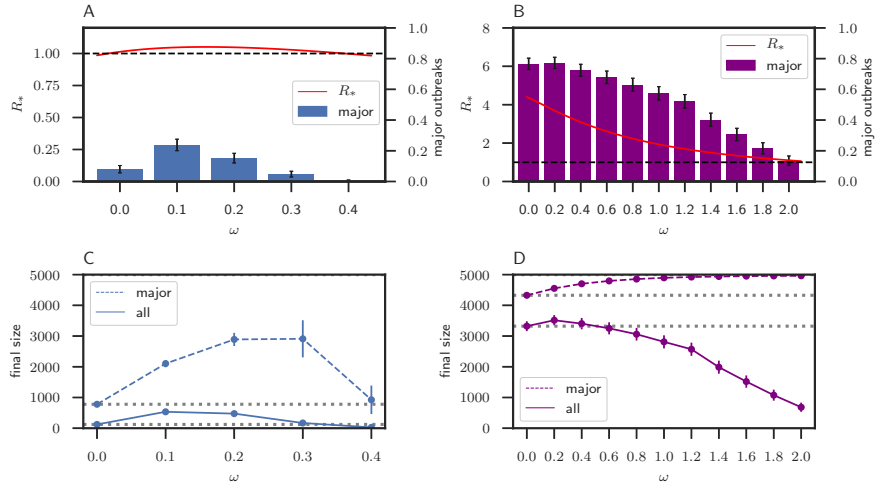}
\caption{The effect of social distancing on the epidemic threshold parameter $R_*$ and the final size. The fraction of epidemics resulting in major outbreaks (with 95\% CI) and $R_*$ for (A) mean infectious period $1/\gamma=100$ time units and two-point degree distribution with $p_0=1/2=p_1$ and (B) mean infectious period $1/\gamma=10$ time units and two-point degree distribution with $p_0=1/2=p_3$. Average final size with (dashed) and without (solid) conditioning on a major outbreak (with $95\%$ CI) corresponding to (C) scenario A (D) scenario B; dotted horizontal lines are for comparison with the size at $\omega=0$. Other parameter values are as follows: cliques have size 3, the population size is 5000 and $\omega_r/\omega_d=9$. Each epidemic is initiated with one randomly chosen infected individual and for each value of $\omega$, 500 epidemics are simulated.}
\label{fig:household}
\end{figure*}

In the clique network individual preventive social distancing can have a negative population-level effect already at the beginning of an epidemic. To demonstrate this we consider $R_*$ for the clique-network model. The clique reproduction number $R_*$ is derived by differentiating between two types of newly infected cliques. A newly infected clique at first consists of one newly infected individual while the remaining clique members are susceptible. The two types are determined by the way the newly infected individual $u_*$ was infected: (1) $u_*$ was infected by a global neighbour (i.e.\ outside his/her own clique) that it had already before the start of the epidemic or (2) $u_*$ was infected by a global neighbour that it acquired through a social distancing event during the epidemic. The clique reproduction number is the dominant eigenvalue of the $2\times2$ matrix $(K_{ij})_{i,j=1,2}$, where $K_{ij}$ is the expected number of cliques of type $j$ generated by one newly infected clique of type $i$. Details of the derivation of the $K_{ij}$ are found in~\textit{SI Section~S2.1}. We find an explicit expression for $R_*$ that we can analyse as a function of social distancing for different degree distributions (see \textit{SI~Section~S2.2}). We illustrate these analytical results with numerical examples in Fig.~\ref{fig:household}.

As can be seen in Fig.~\ref{fig:household}A, $R_*$ can increase as a function of the social distancing rate $\omega$. In particular, social distancing can move the epidemic threshold $R_*$ from below to above its threshold value of one. In other words, individual preventive measures that are beneficial at the individual level can cause a major outbreak to become possible while without the preventive measures this is not possible. However, this depends heavily on the precise network structure. In Fig.~\ref{fig:household}B, the degree distribution is chosen such that $R_*$ decreases for all social distancing rates. See~\textit{SI~Section~S2} for more details and examples of the dependence of $R_*$ on social distancing. Note that $R_*$ will eventually decrease for large enough social distancing rates as can be seen in Fig.~\ref{fig:household}A.

In settings where social distancing pushes $R_*$ from below to above the threshold for an epidemic to occur, the effect of social distancing on the final size is large (Fig.~\ref{fig:household}C). Moreover, even in settings where social distancing reduces $R_*$, the final size can initially increase when social distancing is introduced into the model (Fig.~\ref{fig:household}D).

\subsection{Application to real-world networks}
\begin{figure}[tbhp]
\centering
\includegraphics[width=.4\linewidth]{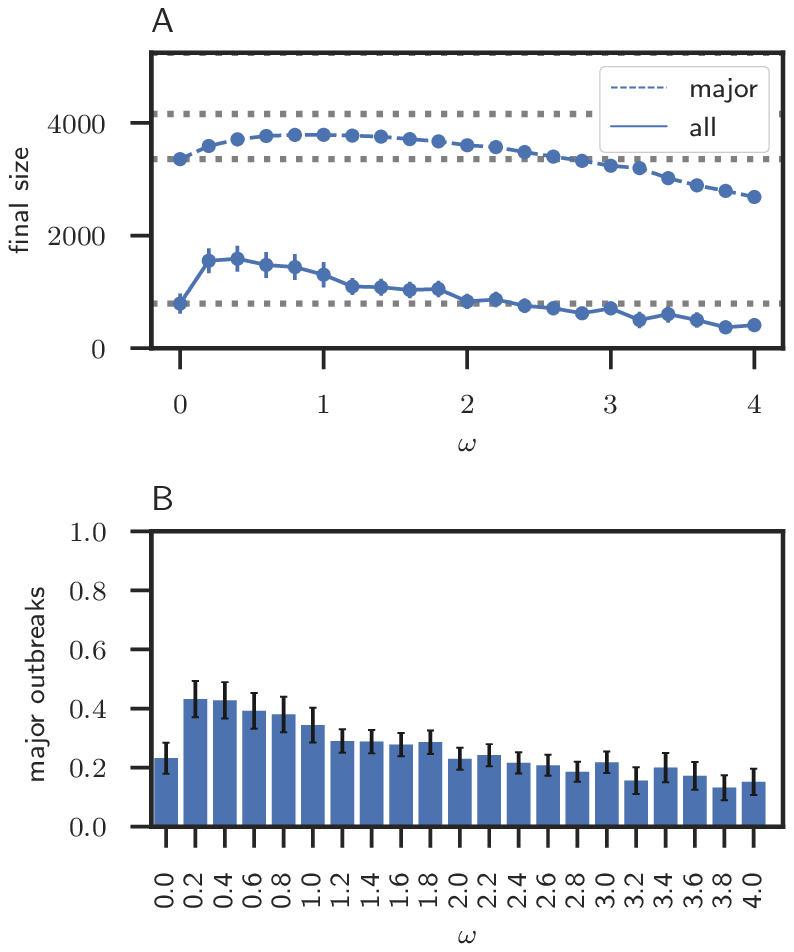}
\caption{Social distancing can increase the final size of the epidemic on real-world networks for large recovery rate. Social distancing in the arXiv collaboration network. (A) The average final size with (dashed) and without (solid) conditioning on a major outbreak (with 95\% CI); dotted horizontal lines are for the size of the giant component (top) and comparison with the size at $\omega=0$ (bottom two). (B) Fraction of all outbreaks that resulted in major outbreaks (with 95\% CI). Model parameter values are: mean infectious period $1/\gamma=2$ time units and social distancing $\omega_r/\omega_d=9$. For each value of $\omega$, 500 epidemics are simulated. The index case is chosen uniformly at random from the sub-population of individuals that has median degree and are part of the largest connected component of the network.}
\label{fig:empirical2}
\end{figure}

We consider two real-world networks: the collaboration network of arXiv General Relativity and social circles from Facebook, taken from~\cite{snapnets}. We simulate SIR epidemics with social distancing on these two real-world networks (see \textit{SI~Section~S3.1} for details). In Fig.~\ref{fig:empirical2} we demonstrate that social distancing can have a negative effect at the population level by increasing the final size in the collaboration network. Further examples with substantially different parameter values but qualitatively the same results are presented in~\textit{SI Fig.~S5}. 

\begin{figure}[tbhp]
\centering
\includegraphics[width=.4\linewidth]{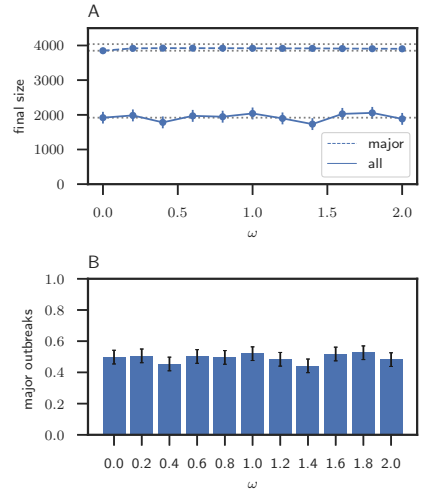}
\caption{Social distancing in the facebook social circles network with randomly chosen index case with median degree. (A) The average final size over all outbreaks (solid) and conditioning on major outbreaks (dashed) (with 95\% CI); dotted horizontal lines are for the size of the network (top) and comparison with the size at $\omega=0$ (bottom two). (B) Fraction of all outbreaks that resulted in major outbreaks (with 95\% CI). Model parameter values are: mean infectious period $1/\gamma=2$ time unites and social distancing $\omega_r/\omega_d=9$. For each value of $\omega$, 500 epidemics are simulated. The index case is randomly chosen from the population that has median degree.}
\label{fig:empirical_facebook}
\end{figure}

The second real-life network that we consider the social distancing epidemic model on is the Facebook social circles in Fig.~\ref{fig:empirical_facebook}. This serves to demonstrate that the precise network structure plays a crucial role for the effect that social distancing can have on the final size. We find that if we restrict to only the major outbreaks, then a modest increase in the final size can be observed when compared to the baseline setting. On the other hand, the average final size is more or less unaffected by social distancing for sufficiently small social distancing rates. This can be explained by the network structure of the underlying population. Since all individuals are part of the same connected component that contains many connections, i.e.\ all individuals are (indirectly) connected to each other, modest social distancing rates will not change the network structure in a way that significantly alters transmission patterns (see \textit{SI Section S3.1} for network summary statistics).

\section{Conclusion and discussion}
In the event of an epidemic outbreak in a population, individuals may take preventive measures by changing their contact patterns. Individuals may try to avoid infection by social distancing from infectious contacts. While this behaviour is always rational at the individual level, it may be harmful at the population level. In fact, preventive social distancing can increase the final epidemic size at the population level and thus have negative effects for the community at large. We demonstrated this counter-intuitive result by means of different epidemic network models, as well as simulating epidemics with social distancing on existing real-world networks. Similar conclusions in terms of behavioural changes at the individual level and its population-level consequences have been drawn in~\cite{Meloni2011,Nicolaides2013} for different behavioural change models. Both~\cite{Meloni2011,Nicolaides2013} considered changes in human mobility patterns in the event of an epidemic and its consequences for the geographical spread. Using a metapopulation model, they illustrated that individual preventive measures in mobility patterns can lead to epidemic spread in new locations, although their invasion thresholds are always increasing~\cite{Meloni2011} or even independent~\cite{Nicolaides2013} of the behavioural changes, which is quite different from the dependence on social distancing of the threshold parameters $R_0$ and $R_*$ in our models.

Whether or not social distancing will actually have negative epidemic outcomes depends strongly on the social network structure of the population. We demonstrated that social distancing can have different effects in the initial stages of the epidemic compared to the overall epidemic outbreak size. We considered the spread of an SIR epidemic on the clique-network model and the configuration network model. We showed that social distancing can have negative effects for the community by (i) increasing the epidemic threshold parameter $R_*$ from below to above the threshold value of one in clique-networks with high clustering and (ii) by increasing the final size. Point (ii) for the final size was shown in (a) configuration networks with heterogeneous degree distribution, (b) clique-networks, and (c) a real-life collaboration network. 


In general, in the baseline setting that an epidemic outbreak may occur when no preventive measures are taken, social distancing can always have beneficial effects provided that the rate of social distancing is sufficiently large (e.g.~Fig.~\ref{fig:empirical2}A). Indeed, sufficiently large social distancing rates can prevent an epidemic from taking off by reducing the epidemic threshold parameter from above to below its threshold value. In such cases, social distancing ensures that only a small number of individuals get infected by the epidemic, while in the baseline setting a significant fraction of the population may be infected. 

Whereas social distancing never increases ones own risk of getting infected in our model, through rewiring, it can increase the risk for other individuals, e.g.\ by connecting to individuals that were previously not (so heavily) exposed to the epidemic. How and whether or not social distancing affects the population-level epidemic outcome depend on a variety of factors. Most notably, the network structure plays an important role (e.g.~Fig.~\ref{fig:household}). Besides the network itself, the infectious disease under consideration is also of importance. For infectious diseases with long mean infectious periods, social distancing can have negative population level epidemic outcomes for a much larger range of social distancing rates compared to infections with shorter mean infectious periods (e.g.~Fig.~\ref{fig:empirical2} and \textit{SI Fig.~S5}). Furthermore, social distancing where the majority of connections are rewired rather than dropped can more easily lead to negative effects at the population level. The same seems to apply when $R_0$ or $R_*$ is high and the community has many individuals with low degrees and/or the community has highly connected cliques. In such cases, rewiring may introduce or increase connections to otherwise relatively isolated individuals. In this way the smaller chance of the individual who takes preventive measures getting infected is outweighed by the increased risk of transmission to a larger part of the population in the event of infection.

Although it is generally recognised that individual preventive measures are often taken once awareness of an epidemic is in place, it is not well understood how to model changes in individual behaviour. Here we considered the effect of social distancing on an epidemic. We modelled this on a contact network by assuming that susceptible individuals distance themselves from infectious contacts, allowing for both dropping of connections and replacement with new contacts in the desire to sustain a certain number of social contacts. Social behaviour is far more complex than our social distancing model, and many behavioural changes will depend on the epidemic and population under consideration (e.g.\ risk perception is an important factor). However, the aim of our paper is to show that rational individual-level preventive measures \emph{can} have counter-intuitive consequences for the population-level. Public health interventions that aim at changing individual behaviour through social distancing could have adverse consequences, for example school closures could reduce social contacts between children in the school classes but may (partly) be replaced by social contacts outside of school. As our results show, it is not necessarily straightforward what effects such behaviour may have at the population level. These findings highlight the importance of modelling individual level behavioural changes in response to an epidemic to understand infectious disease dynamics. 



\subsection*{Acknowledgements}
This work was partially supported by a grant from the Simons Foundation and was carried out as a result of the authors' visit to the Isaac Newton Institute for Mathematical Sciences during the programme Theoretical Foundations for Statistical Network Analysis in 2016 (EPSRC Grant Number EP/K032208/1). T.B.\ and K.Y.L.\ are supported by the Swedish Research Council (VR) Grant Number 2015-050153. This work was also supported by a grant from the Knut and Alice Wallenberg Foundation, which enabled F.B.\ to be a guest professor at the Department of Mathematics, Stockholm University.



\setlength{\bibsep}{2pt}
\bibliographystyle{unsrt}
\bibliography{refs_network}

\end{document}